\documentclass[a4paper]{jpconf}
\usepackage{graphicx}
\begin{document}
\title{A search for extragalactic pulsars in the Local Group galaxies IC 10 and Barnard's Galaxy}

\author{Hind Al Noori $^1$, Mallory S E Roberts $^{1,2}$, David Champion $^3$, Maura McLaughlin $^4$, Scott Ransom $^5$ and Paul S Ray $^6$}
\address{$^1$ New York University Abu Dhabi, Abu Dhabi, UAE}

\address{$^2$ Eureka Scientific, Oakland, CA. USA}

\address{$^3$ Max-Plank-Institut fr Radioastronomie, Bonn, Germany}

\address{$^4$ West Virginia University, Morgantown, WV. USA}

\address{$^5$ National Radio Astronomy Observatory, Charlottesville, VA. USA}

\address{ $^6$ Naval Research Laboratory, Washington, DC. USA}

\ead{haa280@nyu.edu}

\begin{abstract}
As of today, more than 2500 pulsars have been found, nearly all in the Milky Way, with the exception of $\sim$28 pulsars in the Small and Large Magellanic Clouds. However, there have been few published attempts to search for pulsars deeper in our Galactic neighborhood.  Two of the more promising Local Group galaxies are IC 10 and NGC 6822 (also known as Barnard's Galaxy) due to their relatively high star formation rate and their proximity to our galaxy. IC 10 in particular, holds promise as it is the closest starburst galaxy to us and harbors an unusually high number of Wolf-Rayet stars, implying the presence of many neutron stars.  We % searched 
observed IC 10 and NGC 6822 at 820 MHz with the Green Bank Telescope for $\sim$15 and 5 hours respectively, and put a strong upper limit of 0.1 mJy on pulsars in either of the two galaxies.  We also performed %preliminary
single pulse searches of both galaxies with no firm detections.

\end{abstract}

\section{Introduction and Background} 

	Almost 50 years have passed since Jocelyn Bell serendipitously discovered the first pulsar after noticing periodic fluctuations in radio telescope data \cite{bell}.
	% the same year Pacini  predicted that an oblique magnetized rotating star should emit radiation along its magnetic axis \cite{pacini}. This was few years after Hoyle, Nelikar and Wheeler first proposed the presence of electromagnetic emission from rotating neutron stars \cite{hnw}. 
Since then, over 2500 pulsars have been discovered \cite{manchester}. The vast majority of these pulsars have been found in the Milky Way, with only 23 found in the Large Magellanic Cloud (LMC) and 5 in the Small Magellanic Cloud (SMC) \cite{ridley}, both of which are satellites of our galaxy. This work is an attempt to search for radio pulsars that are truly extragalactic, lying beyond the Magellanic Clouds.

%	These objects are of interest for many reasons. Not only are they interesting as astronomical laboratories where magnetic fields impossible to create on earth can be studied, they can also be used to obtain a wide array of valuable information across many domains of physics. Pulsars are incredibly compact objects, second only to black holes, and so their study can provide tests of general relativity as detectors of gravitational waves \cite{array, nanograv}. The precision in pulsar spin periods also makes it possible for them to theoretically be used for spacecraft navigation in deep space, an option that may be particularly favorable for distant missions \cite{spacenav}. Additionally, the process of pulsar searching provides us with a probe for studying the interstellar medium (ISM). 	

	The detection of extragalactic pulsars would open up new avenues in pulsar astronomy. What we can now learn about the ISM from Galactic pulsars, extragalactic pulsars would teach us about the intergalactic medium (IGM) and the ISM in other galaxies, which we currently know little about \cite{meiksin}. Additionally, the study of pulsars in other galaxies would provide us with yet another tool by which we can study stellar evolution in galaxies different from our own. Motivated by these reasons, we present searches of two local group galaxies, IC 10 and NGC 6822 (also known as Barnard's Galaxy).

\subsection{IC 10}  

	IC 10 is the closest starburst galaxy to the Milky Way, and is in fact the only starburst in the Local Group (LG). It lies fairly low in the Galactic plane (\textit{b} $= -3.3^\circ$), making distance measurements particularly difficult. However, most recent measurements have found the galaxy, using tip of the red giant branch and Cepheids to determine distance, to be in the neighborhood of 660$-$740 kpc away \cite{sakai, demers}.  The galaxy has been classified as a blue compact dwarf \cite{richer} and has quite a high rate of star formation (SF), estimated to be around 3-4 times greater than the SF rate of the SMC \cite{leroy}. It is thought to be currently undergoing vigorous neutron star formation resulting from a SF burst $ \sim$6$-$10 Myr ago. Recent research has shown that the younger population of IC 10, the stars resulting from the most recent SF burst, is found near the centroid of the galaxy, while the oldest population ( $>$2 Gyr) is significantly offset from the center \cite{gerbrandt}. IC 10 shares many characteristics with the Magellanic Clouds and the SMC in particular: they both have active SF, are metal poor, and have extended HI envelopes. This makes the SMC a good template for comparison with IC 10, and gives us a good reason to choose the galaxy for this search.
	
	The evolution of this galaxy is of particular interest as it has an extraordinary number of Wolf-Rayet (W-R) stars, with 33 stars confirmed spectroscopically while a total of $ \sim$100 W-R stars are estimated to exist in IC 10 \cite{crowther, masseyholmes}.  It has an anomalously high ratio of carbon-type (WC) to nitrogen-type W-R (WN) stars, twenty times the ratio expected for a galaxy of its metallicity \cite{massey, massey07}, although this ratio could change as more stars are confirmed spectroscopically \cite{massey15}. Although it has half the effective radius of the SMC, the galaxy has an effective W-R density 5 times that of the SMC, similar to the densities found in OB associations, and two times greater than that of any other galaxy in the LG \cite{wilcotsmiller}.

\subsection{Barnard's Galaxy}

	Besides the Magellanic Clouds, Barnard's galaxy is the nearest dwarf irregular galaxy to us, and is a prototypical dwarf irregular galaxy. Its close distance, $\sim$470 kpc, makes it one of the few dwarf irregulars whose ISM and stellar population can be studied in detail \cite{cioni}. This makes it an excellent candidate to search for pulsars, as probing its pulsar population could give us hints that could be extrapolated to other dwarf irregulars farther away. NGC 6822 has had a consistent, albeit modest, rate of SF for the last 400 Myr, but also experienced a recent increase in SF rate that is thought to have occured 100$-$200 Myr ago \cite{gallart}. It has 4 W-R stars, and a W-R density similar to that of the SMC \cite{AM1991}. Due to the galaxy lying in the direction of the Galactic center (\textit{b} $= -18.4^\circ$), it is obscured and has high reddening and absorption, making it hard to detect things such as supernova remnants (SNRs) optically, of which only one has been detected \cite{kong}. 
	
\section{Observations and results}
\subsection{Search Methodology}

	We conducted three observations of IC 10 and two of Barnard's Galaxy using the Green Bank Telescope in West Virginia. The observations were done using the GUPPI backend with 2048 spectral channels, observing at a central frequency of 820MHz and a bandwidth of 200MHz. The IC 10 observations were taken on the 13th, 14th and 17th of October 2009 for 6, 5, and 4.5 hours respectively, and all had a time resolution of 204.6 $\mu$s. The Barnard's galaxy observations were made on the 10th and 17th of October 2014, for approximately 2.5 hours each. We used a better time resolution with NGC 6822, for which we had a sample time of 81.92 $\mu$s. The galaxies are both well contained within the GBT beam.
	
	Although these two galaxies are much farther than the LMC and the SMC (that have distances of 50kpc and 60kpc respectively), where currently the farthest pulsars are known, we still expect to be sensitive to the most luminous pulsars at these greater distances. Assuming that pulsars born in dwarf galaxies are similar in characteristics to the ones in the Milky Way, and that the distances we use are correct, we would optimistically hope to find $\sim$1 in each galaxy. We base this estimate on the luminosities of Milky Way pulsars.

	We used the software package PRESTO  to search each observation up to a dispersion measure (DM) of 3000 pc cm$^{-3}$ \cite{presto}. The NE2001 model for Galactic free electron density predicts a maximum Galactic contribution to the DM of 210 pc cm$^{-3}$ in the direction of IC 10 and DM of 108 pc cm$^{-3}$ in the direction of Barnard's Galaxy \cite{ne2001}. A newer model by Yao, Manchester and Wang designed to predict distances and DMs of extragalactic pulsars predicts a DM of 289 pc cm$^{-3}$ in the direction of IC 10 and 78 pc cm$^{-3}$ in the direction of NGC 6822. Both of these values are predominantly due to galactic effects, as the model predicts negligible DM coming from the IGM \cite{yao}. We chose to search up to a DM of 3000 pc cm$^{-3}$, several multiples of the predicted electron density to be safe, since very little is known about the electron densities of the IGM and the ISMs of each galaxy, and to allow for detecting fast radio bursts (FRBs) which could serendipitously occur in the direction of these galaxies. We performed an acceleration search up to a zmax of 300 to search for pulsars in tight binaries.
	
	As we have multiple observations of each source, we are able to check the legitimacy of our pulsar candidates by making sure they appear in all data sets. In addition to regular pulsed emission, we search for giant single pulses, which McLaughlin and Cordes predict should be detectable in each of the LG galaxies if they harbor a Crab-like pulsar \cite{giant}. Giant single pulses, such as the ones emitted by the Crab pulsar \cite{crab} can be can be over a hundred times brighter than the regular pulsed emission from pulsars \cite{cognard}. We expect to be much more sensitive to giant pulses than we are to regular pulsations.
 \subsection{Results}

Unfortunately, no strong pulsar candidates were detected, and none of the marginal candidates were seen in more than one pointing. We use the radiometer equation to calculate upper limits of 0.05 mJy and 0.07 mJy for IC 10 and NGC 6822 respectively, assuming a pulse width of $\sim$10\% of the period.
 %we have thus far been unable to confirm any pulsars using the aforementioned check.Despite the plethora of candidates we obtain, none seem particularly promising or worthy of follow-up. One of the difficulties of a search such as this, is that it has never been done before, and so it is difficult to discern what an acceptable pulsar candidate might be.
 A preliminary search for single giant pulses was done for each of the two galaxies, with no obvious single pulse detections. However, a more thorough single pulse search is under way.
 
\section{Discussion}

%Considering our failure to detect any strong pulsar candidates in the two galaxies, it would be constructive to address why our search yielded no new pulsars. A simple way to answer this question would be that our instruments are not yet sensitive enough for pulsar detections at these distances. For example, there could be yet unknown effects of the IGM that result in a net loss of sensitivity. Another possibility, is that we simply need to do some more signal processing to the data in order to improve the signal-to-noise ratio and be able to detect the distant pulsars. But if we assume that we have sufficient sensitivity and don't need to do any further processing, why have we not detected any pulsars?

 \subsection{The case of IC 10}
 
From the null result of our search, we may say that there are not many bright pulsars in IC 10. Much effort has gone into understanding the SF rate of IC 10; the literature shows that SF in the galaxy has happened in bursts rather than continuously \cite{sanna}, and that the most recent SF burst occured within the past 10 Myr, likely to have been as recently as 3$-$4 Myr ago \cite{magrini, vacca}. Such recent SF may explain why IC 10 is home to so many W-R stars and yet have few bright pulsars. Antoniou et al. argue that a 10 Myr old SF burst is too young to make any pulsars, yet can produce black holes \cite{antoniou}. Coincidentally, IC 10 is home to the most massive known stellar mass black hole IC 10 X-1 \cite{prestwich, silverman}.
 
 IC 10 has several large HI holes and a large synchrotron bubble that were thought to have resulted from multiple SNRs \cite{skillman}. However it appears more likely that the synchrotron bubble is actually a hypernova remnant, from what could plausibly be the event that created IC 10 X-1 \cite{moiseev}. This further supports the hypothesis that the recent SF burst is too recent to have resulted in any pulsars. Besides the synchrotron bubble, there are no other SNRs in IC 10.
 
 Another possibility is that IC 10's SF rate has been grossly overestimated \cite{leroy}. If a top-heavy initial mass function (IMF) applies to IC 10, that could explain its high density of W-R stars, but could also mean that its SF rate could be as much as 10 times smaller than we expect. 
 
 \subsection {The case of Barnard's Galaxy}
 
 Unlike IC 10, NGC 6822 has had modest SF for the past 400 Myr, which has had an almost two-fold increase over the last 100$-$200 Myr \cite{wyder}. There is also evidence that SF rate has gone down around 10 Myr ago \cite{gallart}. Since the lifetimes of most longer-period pulsars tend to be closer to 10 Myr than 100Myr, it is likely that the some of the pulsars born during the peak of Barnard's galaxy's SF peak are no longer energetic enough to emit pulses that would be detectable at this distance.
 
 \ack
This research was carried out on the High Performance Computing resources at New York University Abu Dhabi. Many thanks to Jorge Naranjo for HPC support. Contributions to this work by PSR at NRL are supported by the Chief of Naval Research (CNR). The National Radio Astronomy Observatory is a facility of the National Science Foundation operated under cooperative agreement by Associated Universities, Inc.

\end{document}